\documentstyle[11pt,newpasp,twoside,epsf]{article}
\def\edcomment#1{\iffalse\marginpar{\raggedright\sl#1\/}\else\relax\fi}
\reversemarginpar

\begin{document}
\title{The Cepheid Instability Strip in the GAIA Era}
\author{Giuseppe Bono}
\affil{INAF - Rome Astronomical Observatory, Via Frascati 33, 
00040 Monte Porzio Catone, Italy; bono@mporzio.astro.it}

\begin{abstract}
We present recent results concerning distance determinations based 
on the two most popular primary distance indicators, namely classical 
Cepheids and RR Lyrae. We discuss the problems affecting the Cepheid
distance scale, and in particular the dependence of fundamental 
Period-Luminosity (PL) and Period-Luminosity-Color (PLC) relations 
on the metal content. The key advantages in 
using the K-band PL relation of RR Lyrae stars when compared with the 
$M_V$ vs [Fe/H] are also presented. We outline the impact that GAIA's 
spectroscopic measurements will have not only on the distance scale 
but also to constrain the gradients of metals and $\alpha-elements$ 
(see the paper by Thevenin in these proceedings) across the disc and 
the halo as well as current theoretical predictions concerning Galactic 
models.  
\end{abstract}

\section{Introduction}

The bulk of stars with spectral types ranging from late A to late G are 
pulsating variables located inside the so-called Cepheid instability 
strip. When moving from brighter to fainter objects inside this strip 
we find Classical Cepheids and RR Lyrae stars. These objects are the 
prototypes of young, intermediate-mass and old, low-mass standard 
candles. Owing to these intrinsic features and to the fact that they 
can be easily identified due to their large luminosity variation they 
are robust stellar tracers. Between these two groups of variables 
are located Type II Cepheids. Their periods are similar to classical 
Cepheids but they are old, low-mass stars in the Asymptotic-Giant-Branch 
phase (double-shell burning). This evolutionary phase is substantially 
shorter when compared to RR Lyrae and Cepheids. 

Toward fainter magnitudes we 
find the $\delta$ Scuti stars (Breger 2002) and the Oscillating Blue 
Stragglers (Bono et al. 2002). These objects are intermediate-mass stars  
burning Hydrogen in the core or in a thick-shell, i.e. evolved off the 
main sequence. Obviously these variables outnumber the previous ones, 
but the luminosity amplitudes are smaller and the mode identification 
is still debated in the literature.  
Stars with spectral types ranging from late K to M and low surface 
gravities are once again pulsating variables located in the Mira 
instability strip. The physical mechanism driving the pulsation 
instability is the same, but the pulsation properties are substantially 
different. The most common variables located inside this strip 
are the semiregular variables and the Long-Period-Variables (see the 
paper by Feast in these proceedings).  

Instruments on board of GAIA have been optimized to supply accurate 
trigonometric parallaxes and physical parameters (chemical composition, 
radial velocity, effective temperature, surface gravity, reddening, 
binarity) for stars with spectral types ranging from F to K. This 
means that photometric and spectroscopic data collected by GAIA will 
have a fundamental role to improve current knowledge on pulsating 
variables located inside the Cepheid and the Mira instability strips.    

The microlensing experiments (EROS, OGLE, MACHO, PLANET), aimed at 
the detection of baryonic dark matter have already provided a large 
amount of photometric data in the direction of the Galactic bulge. 
These data substantially increased the number of variables for which 
are available accurate estimates of periods, mean magnitudes and 
colors. However, the observables that GAIA plan to measure are mandatory
(see the paper by Perryman in these proceedings) to constrain the 
accuracy of theoretical predictions concerning evolutionary and 
pulsation properties of Galactic stars, the formation and evolution 
of the Galaxy and its interaction  with nearby dwarf galaxies as well 
as chemical evolution models. 

In the following we discuss the role that GAIA spectroscopy will have 
on Cepheid and RR Lyrae distance scales as well as on the use of these 
objects as stellar tracers. Finally, we briefly outline the impact 
that GAIA will have on outer disc and halo stellar populations.    
 
\section{Classical Cepheids}

The Cepheid distance scale is the crossroad for the calibration of 
secondary distance indicators, and in turn for estimating the Hubble 
constant $H_0$. The strengths and the weaknesses in using the 
PLC relation, that supplies individual 
Cepheid distances, or the PL relation, that 
supplies ensemble distances, have been widely discussed in the 
recent literature (Sandage et al. 1999; Tanvir 1999). Pros and cons 
in using optical, near-infrared (NIR), or Wesenheit magnitudes 
(W=V-2.45(V-I)) reddening free magnitudes) have also been lively 
debated during the last few years (Feast 1999; Freedman et al. 2001;
Bono 2003). However, the critical issue concerning Cepheid distances 
is to assess on a firm basis whether the zero-point and the slope of 
PL and PLC relations do depend on the metal content. During the last
few years the number of theoretical and empirical investigations  
focused on this problem are countless. Empirical findings seem to 
suggest that the PL relation in the optical bands presents a mild 
dependence on the metallicity. In particular, metal-rich Cepheids 
at fixed period appear {\em brighter} than metal-poor ones 
(Sasselov et al. 1997; Kennicutt et al. 1998; Sandage et al. 1999; 
Macri et al. 2001; Fouqu\'e et al. 2003). 

On the other hand, hydrodynamical envelope models that account for 
the coupling between pulsation and convection predict that metal-rich 
Cepheids, at fixed period, are {\em fainter} than metal-poor ones
(Bono et al. 1999a,b; Alibert et al. 1999). This means that we are 
facing with a substantial discrepancy between theoretical predictions 
and empirical data. Current scenario is further {\em jazzed up} by 
multi-band analyses of Magellanic Cloud (MC) Cepheids that support 
the theoretical 
sign (Groenewegen \& Oudmaijer 2000; Groenewegen 2000). At the same 
time, recent empirical investigations based on accurate individual 
distance and reddening determinations of Galactic Cepheids appear
to suggest that both the zero-point and the slope of the PL relation 
in the optical bands do depend on the metal content (Saha et al. 2003;
Tammann \& Sandage 2003). {\em Paucis verbis} the Cepheid PL and PLC 
relations in the optical bands are not universal. This means that to 
estimate the distance of external galaxies are necessary PL and PLC 
relations based on Cepheids that present the same mean metallicity.    
As a consequence, Galactic Cepheids might be crucial to improve the 
intrinsic accuracy of distance determinations, since the metallicity 
distribution of these objects, in contrast with Magellanic ones, is 
quite similar to the mean metal abundance of external spiral galaxies 
where the two HST key projects identified Cepheids (Freedman et al. 2001; 
Saha et al. 2001). 
  
However, it is worth mentioning that theoretical and empirical 
results do suggest that the PL and the PLC relation of First 
Overtone (FO) Cepheids marginally depends on metal content. 
In a recent investigation Bono et al. (2002) found that 
predicted and empirical Wesenheit function as well as $PL_K$ 
relations provide quite similar mean distances to the MCs. 
This finding seems quite promising, since these 
distance evaluations are marginally affected by systematic 
uncertainties. In fact, K-band and Wesenheit magnitudes are marginally 
affected by uncertainties on reddening corrections and presents 
a mild dependence on metallicity (Bono et al. 1999b). The width in 
temperature of FO Cepheids is roughly a factor of two narrower than 
for fundamental (F) Cepheids. Therefore distances based on former 
variables are marginally affected by the intrinsic spread of the 
PL relation typical of the latter ones. Moreover, the period range 
covered by FOs is $\approx$ 1 dex shorter than for F Cepheids, thus 
FO PL and PLC relations are hardly affected by changes in the slope
when moving from long to short-period variables (Bauer et al. 1999;
Bono, Caputo, \& Marconi 2001).   

{\em Why GAIA might play a crucial role to improve the Cepheid 
distance scale?}  

To nail down the systematic uncertainties affecting the Cepheid 
distance scale are required accurate trigonometric parallax 
measurements for a sizable sample of Galactic Cepheids. However,  
according to the above discussion accurate and homogeneous 
{\em measurements} of Cepheid metal abundances are mandatory 
to properly address the problem. The most recent spectroscopic 
investigation concerning the chemical composition of Galactic 
Cepheids do rely on a sample of roughly 100 objects (Andrievsky et al. 
2002b, and references therein). According to these authors the 
metallicity distribution across the Galactic disk might be 
split intro three different zones: 

{\em i)} the inner region, ranging 
from $\approx 4$ to $\approx 7$ kpc for which they found a  metallicity 
gradient of $d [Fe/H]/dR_G\approx-0.13\pm0.03\; dex\; kpc^{-1}$; 

{\em ii)} 
the central region, ranging from $\approx 7$ to $\approx 10$ kpc, for 
which the gradient is flatter and equal to $\approx-0.02\pm0.01\; dex 
\;kpc^{-1}$; 

{\em iii)} while in a small portion of the outer disk (toward 
the Galactic anticenter), ranging from $\approx 10$ to $\approx 12$ 
kpc, they found a gradient of $\approx-0.06\pm0.01\; dex \;kpc^{-1}$. 
Moreover and even more importantly, Andrievsky et al. found a discontinuity 
in the metallicity distribution located at roughly 10 kpc from the Galactic 
center. These findings somehow supports the results obtained 
by Twarog et al. (1997) on the basis on photometric (Stroemgren) indicators 
and by Caputo et al. (2001) on the basis of multiband (Johnson) 
Cepheid pulsation relations. However, no firm conclusion can be 
drawn concerning the 
metallicity gradient, since current photometric and spectroscopic data 
for Cepheids located in the outer disc ($d \ge 12$ kpc) are scanty. 
  
\begin{figure}  
\plotone{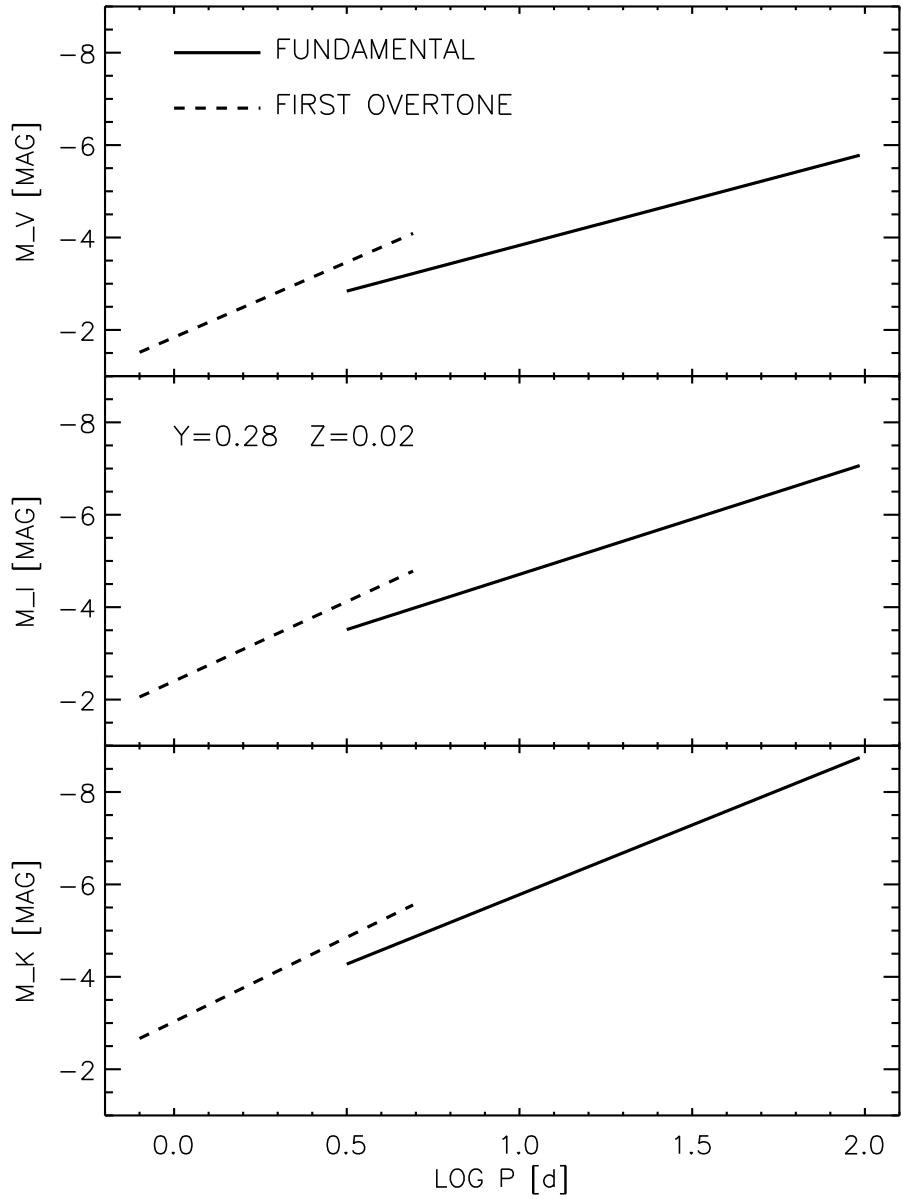}
\vspace*{1.0truecm}\\  
\caption{Theoretical PL relations in V (top), I (middle), and K (bottom) 
band at solar chemical composition (Bono et al. 1999b). Solid and dotted 
lines display F and FO PL relations.} 
\end{figure}

It goes without saying that homogeneous high resolution spectra for a 
complete sample of Galactic Cepheids together with accurate distances 
might be the {\em Panacea} not only for the Cepheid distance scale but 
also to constrain dynamical models of the galactic disc as well as 
its chemical evolution. Note that the occurrence of shallow 
metallicity gradients of iron and of iron-group elements might be 
caused, according to current predictions by gas infall from the 
halo, by gas viscosity in the disc, or by a central bar structure 
(Portinari \& Chiosi 2000; Andrievsky et al. 2000a). This scientific 
goal can be easily reached by GAIA, since a spectrograph with  
$R\approx15000$ should supply accurate spectra ($S/N\ge 50$) 
Data plotted in Fig. 1 show that short-period, Galactic FO 
Cepheids present and absolute I magnitude ranging from -1.5 to -2.0. 
Therefore if we assume that the outskirt of the Galactic disc is 
roughly located between 20 and 25 kpc ($DM\approx 17.0$ mag) and a
mean reddening that ranges from E(B-V)=0.5 to 1.0, then the apparent 
magnitude of fainter Cepheids should range from $I\sim16$ to 
$I\sim17$ mag. This means that a substantial fraction of Galactic 
Cepheids are brighter than the limiting magnitude ($V\sim17.5$ mag) 
for which GAIA will supply accurate chemical compositions and radial 
velocities (see the papers by Munari and Zwitter in these proceedings). 

Note that GAIA spectra will be collected at least over 
three consecutive years, therefore they can also be adopted to 
identify a substantial fraction of binary Cepheids (see the paper by 
Szabados in these proceedings).  
Finally, we mention that this project can be hardly accomplished 
with current generation of multi-fiber, wide field of view spectrographs
such as FLAMES/GIRAFFE@VLT, since the spatial density of Cepheids is 
too low.

\section{RR Lyrae variables}

RR Lyrae stars are very useful objects, since they are robust, low-mass 
standard candles and robust tracers of old stellar populations. 
They are ubiquitous across the Galactic spheroid and thanks to their 
pulsation properties (peculiar shape of the light curves, narrow period 
range), they can also be easily identified in Local Group (LG) galaxies. 
Although, RR Lyrae stars present several advantages, distance estimates 
based on different calibrations 
(Baade-Wesselink method, HB models) of the $M_V$ vs [Fe/H] relation 
taken at face value present a difference that is systematically larger 
than the empirical uncertainties (Walker 2000,2003; Cacciari 2003). 
This might indicates that current 
RR Lyrae distance determinations are still affected by systematics.  
Fortunately enough, empirical evidence dating back to Longmore et al. 
(1990) suggest that RR Lyrae stars do obey to a well-defined PL relation
in the K-band ($PL_K$). The use of this relation might overcome some of the 
problems affecting the RR Lyrae distance scale, since the $PL_K$ relation 
is marginally affected by evolutionary effects as well as by the spread 
in stellar mass inside the instability strip. This finding was further 
strengthened by a recent theoretical investigation (Bono et al. 2001) 
suggesting that RR Lyrae obey to a very tight $PLZ_K$ relation 
connecting the period, the luminosity, the K-band absolute magnitude, 
and the metallicity. This approach seems very promising and should 
allow us to supply during the next few years an accurate calibration 
of the $M_V$ vs [Fe/H] relation over the metallicity range covered by 
RR Lyrae ($-2.2 \le [Fe/H] \le 0$).  

During the last few years RR Lyrae are becoming very popular, since the  
Sloan Digital Sky Survey (SDSS) detected an overdensity of candidate 
RR Lyrae and of A-type stars located approximately 50 kpc from the 
Galactic center. According to this empirical evidence Ivezic et al. (2000) 
and Yanny et al. (2000) suggested that such a clump is the northern tidal 
stream left over by the Sagittarius dwarf spheroidal (dSph). 
Independent observational 
(Ibata et al. 2001; Martinez-Delgado et al. 2001; Vivas et al. 2001) and 
theoretical (Helmi \& White 1999; Helmi 2002) investigations support this 
hypothesis. The observations of such extra-tidal stellar remnants in 
dSph resembles the tidal debris recently detected in a large number of 
Galactic Globular Clusters (GGCs, Leon et al. 2000; Odenkirchen et al. 2002). 
On the other hand, dSph galaxies apparently 
host large amounts of Dark Matter (DM), and indeed the mass-to-light 
ratio in these systems range from $(M/L)_V\sim 5$ (Fornax) to $\sim100$ 
(Ursa Minor), whereas in GGCs the M/L ratio is $\approx1-2$. As a 
consequence, the study of the radial distribution of RR Lyrae can 
supply tight constraints on the tidal interaction that 
these interesting systems undergo with the Milky Way. 

On the basis of V-band time series data that cover a large 
sky area (100 deg$^2$)
Vivas et al. (2000) identified and measured the mean magnitude of
148 RR Lyrae stars and more than 50\% of this sample belong to the
clump identified by Ivezic et al. (2000). These data provided the 
first firm evidence that the Galactic halo does not show smooth 
contours in density. In fact, they also detected two smaller 
overdensities in the halo one of which located at R$\approx$ 17 kpc 
seems related to the GGC Palomar 5, while the other is located at 
R$\approx$ 16 kpc. 
 
\begin{figure}  
\plotone{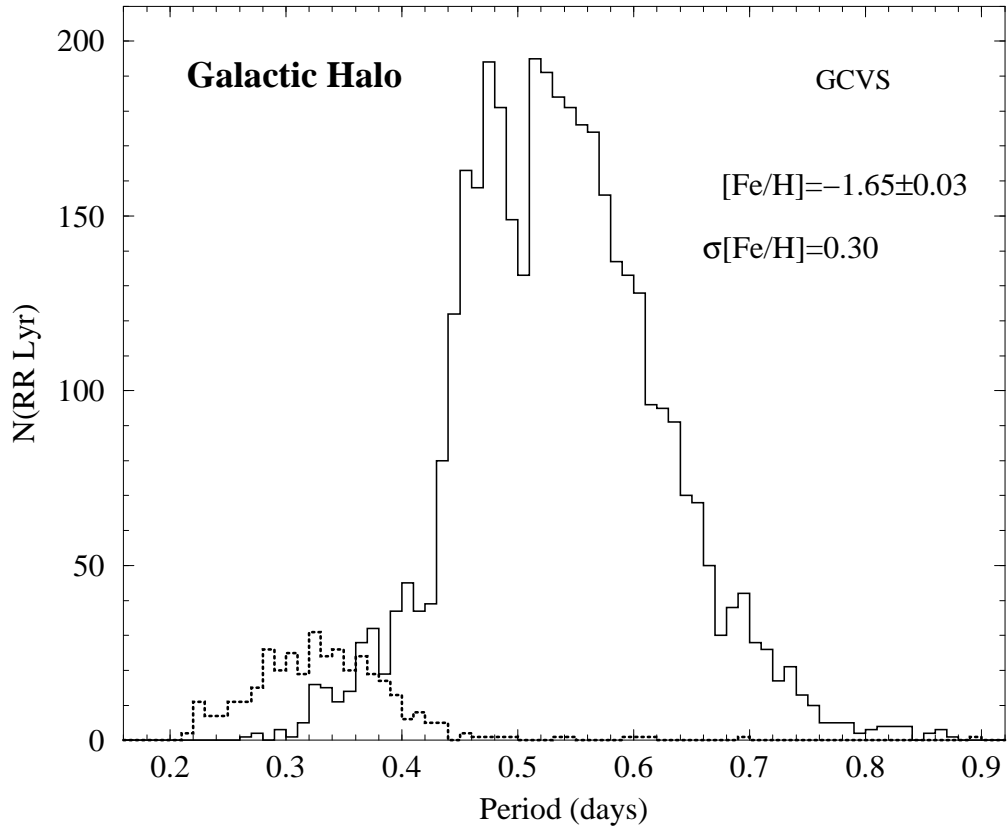}
\caption{Period distribution of RR Lyrae stars in the Galactic halo 
according to the General Catalog of Variable Stars (GCVS). 
Solid and dashed line refer to fundamental and first overtone variables,
respectively. The mean metallicity and the intrinsic spread in metallicity 
are labeled (Suntzeff et al. 1994).} 
\end{figure}

to pin point peculiar radial distributions 
It is worth stressing, that a substantial improvement 
in the intrinsic accuracy of the $M_V vs [Fe/H]$ and of the $PLZ_K$ 
relation does not allow us to use RR Lyrae in the Galactic halo 
to constrain the dynamical interaction of dwarf galaxies and GGCs 
with the Milk Way. The detections of peculiar radial distributions
is hampered by the limited number of RR Lyrae stars for which are 
available accurate spectroscopic measurements of radial velocities 
and chemical compositions (Suntzeff et al. 1994; Layden et al. 1996; 
Dambis \& Rastorguev 2001). Moreover, current 
sample of RR Lyrae in the Galactic halo might also be affected by a 
selection bias. Data plotted in Fig. 2 show that RR Lyrae in the 
halo might be peculiar. The period distribution and the mean period 
of fundamental pulsators ($<P_{ab}>\approx0.539$ d)  mimic the behavior 
of Oosterhoff type I clusters (see figure 3 in Clement et al. 
2001)\footnote{GGCs that host a good sample of RR Lyrae stars are 
classified Oosterhoff type I clusters if the mean fundamental period 
is roughly equal to $<P_{ab}>\approx0.55$ and the ratio between first 
overtones and the total number of RR Lyrae, $N_{FOs}/N_{RR}$ is roughly 
equal to 0.2. A GGC is classified as an Oosterhoff type II if 
$<P_{ab}>\approx0.64$ and $N_{FOs}/N_{RR}\approx0.5$.}. However, the 
number of FO RR Lyrae in the halo is quite small ($N_{FOs}/N_{RR}< 0.1$) 
and the period distribution of fundamental pulsators does show a gap at 
$P\approx0.5$ that is not present among RR Lyrae in the Galactic bulge 
and in GGCs (Bono et al. 2003). At present, it is not clear whether 
these peculiarities are intrinsic or due to a selection bias.  

Once again GAIA might play a crucial role to improve current empirical 
scenario. The unprecedented opportunity to supply accurate high 
resolution spectra down to a limit magnitude of $V\sim17.5$ mag will 
allow us to trace the pulsation properties of RR Lyrae over a substantial 
portion of the halo. The use of metallicity indicators based on the GAIA 
multi-band photometric system should also allow us to extend the 
spectroscopic analysis from the bulge to the outermost regions of the 
halo ($DM\approx19-20$). 

Finally, it is worth mentioning that we still lack an empirical 
estimate of the dynamical mass of a Horizontal-Branch (HB) star, since 
no binary system has been detected that include one of these objects.   
The detection of a few of these systems would be of paramount relevance 
to constrain the input physics (equation of state, opacities, nuclear 
burning rates) of evolutionary and pulsation models, and in turn 
to properly address long-standing stellar astrophysical problems 
such as the second parameter problem and the HB morphology (Castellani 
1999).  

\section{Final Remarks}

The compelling results obtained by the two HST key projects concerning 
the estimate of the Hubble constant contributed to the diffuse believe 
that problems affecting the calibration of both primary and secondary 
distance indicators have been settled. Recent findings concerning the   
dependence of the Cepheid PL relation on the metal abundance, as well 
as the nonlinearity of the RR Lyrae $M_V$ vs [Fe/H] relation 
(Caputo et al. 2000) cast some doubts on this view.   
The mismatch between distance determinations based on different 
standard candles 
further strengthens this working hypothesis. Data listed in Table 1 
clearly show that distance determinations to the Coma cluster based 
on different zero-points and secondary indicators range from 34.64 
to 35.29 mag, while $H_0$ evaluations range from 60 to 84 
$km\,s^{-1} Mpc^{-1}$. Note that the Coma cluster will play a 
fundamental role to improve the accuracy of the Hubble constant, 
since it is the nearest galaxy cluster not affected by local motions.

\begin{table}
\caption{Compilation of distance determinations to the Coma cluster, 
according to different primary and secondary distance indicators.\\}
\begin{tabular}{lccccc}
\tableline 
Method$^a$& Target(s)    & ZP$^b$   & $(m-M)_0^c$    & $H_0^d$  & Ref.$^e$\\ 
\tableline 
        &                &          &                &          &       \\ 
 GCLF   & NGC4874/IC4051 & Virgo$^f$& $35.05\pm0.12$ &$69\pm9$ &   1   \\ 
SBF$_K$ & NGC4874        & Ceph$^g$ & $34.99\pm0.21$ &$71\pm8$ &   2   \\ 
SBF$_{K^{\prime}}$&NGC4889&Ceph$^h$ & $34.64\pm0.25$ &$85\pm10$&   3   \\ 
SBF$_I$ & NGC4881        & Leo-I$^i$& $35.04\pm0.31$ &$71\pm11$&   4   \\ 
SBF$_I$ & NGC4881        & Ceph$^j$ & $35.05\pm0.53$ &$73\pm19$&   5   \\
TF$_H$  &20 galaxies     & Ceph$^h$ & $34.94\pm0.13$ &$73\pm4$ &   6   \\ 
TF$_I$  & \ldots         & Ceph$^j$ & $34.66$        &$84\pm13$/$86\pm14$ & 5 \\
FP$_I$  & 81 galaxies    & Ceph$^j$ & $34.67\pm0.15$ &$83\pm6$/$86\pm6$ &   5 \\
$D_n-\sigma$& 81 galaxies& Ceph$^j$ & $34.89\pm0.16$ &$75\pm5$/$78\pm5$ &   7 \\
$D_n-\sigma$(K)& 24 galaxies& Leo-I$^l$& $34.90\pm0.14$ &$75\pm6$         &   8  \\
SNIa        &  5         & Virgo$^m$& $35.05\pm0.49$ &$70\pm15$         &   9 \\
VM$^n$&  \ldots          & Virgo$^n$& $35.29\pm0.11$ &$60\pm6$         &   10 \\
        &                &          &                &          &       \\ 
\tableline 
        &                &          &                &          &       \\ 
\end{tabular}

$^a$ Globular Cluster Luminosity Function (GCLF); Surface Brightness
Fluctuation (SBF); Tully-Fisher (TF) relation; Fundamental Plane (FP); 
$D_n-\sigma$ or Faber-Jackson relation; Supernovae type Ia (SNIa).  
$^b$ Zero-point. 
$^c$ True distance modulus and relative error as given by authors. 
$^d$ Hubble constant ($km\,s^{-1} Mpc^{-1}$) and relative error as 
given by authors .
$^e$ References: 1) Kavelaars et al. (2000);  2) Liu \& Graham (2001);  
3) Jensen et al. (1999); 4) Thomsen et al. (1997); 5) Freedman et al. (2001);
6) Watanabe et al. (2001); 7) Kelson et al. (2000); 8) Gregg (1997);   
9) Capaccioli et al. (1990); Tammann et al. (1999).   
$^f$ Weighted-average true distance modulus based on Cepheids, TRGB, PNLF, and 
SBF, $\mu_0(Virgo)=30.99\pm0.03$. They adopted a recession velocity of 
$V_r\approx7100\pm200\,km\,s^{-1}$. 
$^g$ Six nearby spiral galaxies for which are available HST Cepheid 
distances ($V_r\approx7186\pm428\,km\,s^{-1}$ by Han \& Mould 1992, 
hereinafter HM92). 
$^h$ Cepheid distances to M31 and Virgo Cluster ($V_r$ by HM92). 
$^i$ Average SBF distance to NGC3379 in the Leo-I group based on 
Cepheids ($V_r$ by HM92). 
$^j$ Revised Cepheid distances to Leo-I group, Virgo and Fornax clusters
(Key Project, the adopted $V_r$ values are 7143 and 7392 $km\,s^{-1}$). 
$^k$ Twelve nearby spiral galaxies for which are available HST Cepheid 
distances ($V_r\approx 7143\,km\,s^{-1}$). 
$^l$ Unweighted-average true distance modulus based on Cepheids, TRGB, PNLF, 
and SBF, $\mu_0(Leo-I)=30.17\pm0.01$ ($V_r\approx 7200\pm300\,km\,s^{-1}$).  
$^m$ Maximum-magnitudes vs rate-of-decline for Novae in M31 
($\mu_0=24.30\pm0.20$) and Virgo ($\mu_0=31.30\pm0.40$, 
$V_r\approx 7130\pm200\,km\,s^{-1}$).  
$^n$ They adopted various (6) secondary methods (Jerjen \& Tammann 1993). 
The ZP is based on the Cepheid distance to the Virgo cluster 
($\mu_0=31.60\pm0.09$). 
\end{table}

The above discussion brings forward the evidence that distance 
indicators might require new detailed empirical and theoretical 
investigations to nail down the deceptive errors affecting 
current distance determinations. In the near future different 
roots might shed new lights on this long-standing problem.  

{\em i)} The use of the white-light interferometer, FGS3, on board 
of HST recently provided a new accurate estimate of the trigonometric 
parallax of $\delta$ Cephei (Benedict et al. 2002), and in turn a new 
calibration of the PL relation. This instrument during the next few 
years might supply accurate geometric distances for a handful of nearby 
Cepheids. This means that the zero-point {\em and the slope} of 
both the PL and the PLC relations can be improved.  

{\em ii)} The new CCD camera (ACS) on board of HST should allow the 
detection of FO Cepheids in external galaxies where F Cepheids have 
already been measured. This instrument could supply the unique 
opportunity to cross-check independent distance determinations based 
on the same group of variable stars.   

{\em iii)} During the next few years ground-based survey telescopes 
aimed at detecting near Earth asteroids, such as 
{\em Pan-STARRS} (Kaiser 2002)\footnote{For further information visit 
http://www.ifa.hawaii.edu/\~kaiser/pan-starrs/pressrelease/} 
and {\em LSST}\footnote{For further information visit 
http://www.lsst.org/lsst\_home.html} will be equipped with detectors 
that cover a sky area ranging from one to several square degrees.  
Therefore a detailed sampling of stellar populations down to 
$V\sim24-27$ mag might be accomplished in the near future. 
The same outcome applies for the wide field imagers that are already 
available on telescope of the 8m class such as SUPRIME@SUBARU or will 
become available in a few years such as LBC@LBT. 
The new multi-band time series data will allow a complete census of 
RR Lyrae and Cepheids belonging to the Galaxy as well as to LG 
galaxies.  
 
In this possible scenario, GAIA gives the unprecedent opportunity 
to supply accurate trigonometric parallaxes, as well as accurate  
measurements of radial velocities, and chemical compositions for a 
large amount of Galactic stars. During the next ten years ground-based 
telescopes of the 8m class equipped with multi-object spectrographs 
will supply accurate estimates of stellar parameters for stellar 
populations in stellar systems such as globular clusters and nearby 
dwarf galaxies. However, the stellar density in the outer disc as 
well as in the halo is too low to be interesting targets for these 
instruments. 

Finally, we mention that the selection of the GAIA photometric bands 
is crucial to improve the accuracy of stellar parameters we plan to 
supply. The estimates of stellar abundances strongly depend on the 
accuracy of effective temperature and surface gravity. Moreover 
and even more importantly, the calibration of new metallicity and 
reddening indicators are two outstanding legacies we are looking 
for from the GAIA mission.

I am indebted to my collaborators for helpful discussions and suggestions. 
This work was supported by MIUR/Cofin~2001 under the project: "Origin and 
Evolution of Stellar Populations in the Galactic Spheroid".



\begin{references}

\reference Alibert, Y., Baraffe, I., Hauschildt, P., \& Allard, F. 1999, 
A\&A, 344, 551  
\reference Andrievsky, S.M. et al. 2002a, A\&A, 381, 32 
\reference Andrievsky, S.M., Kovtyukh, V.V., Luck, R.E., Lepine, J.R.D., 
Maciel, W.J., \& Beletsky, Y.V.  2002b, A\&A, 392, 491   
\reference Bauer, F., et al. 1999, A\&A, 348, 175  
\reference Benedict, G. F. et al. 2002a, AJ, 123, 473  
\reference Benedict, G. F. et al. 2002b, AJ, 124, 1695 
\reference Bono, G. 2003, in Hubble's Science Legacy: Future 
Optical-Ultraviolet Astronomy from Space", ed. K.R. Sembach, J.C. Blades, 
G.D. Illingworth, \& R.C. Kennicutt, (San Francisco: ASP), astro-ph/0210068  
\reference Bono, G., Caputo, F., Castellani, V., \& Marconi, M. 1999b, ApJ, 
512, 711 
\reference Bono, G., Caputo, F., Castellani, V., Marconi, M., \& 
Storm, J. 2001, MNRAS, 326, 1183  
\reference Bono, G., Caputo, F., \& Marconi, M. 2001, MNRAS, 325, 1353 
\reference Bono, G., Caputo, F., Marconi, M., \& Santolamazza, P. 2002, 
in Observational Aspects of Pulsating B- and A Stars, ed. C. Sterken \& 
D.W. Kurtz (San Francisco: ASP), 249   
\reference Bono, G., Groenewegen, M. A. T., Marconi, M., Caputo, F. 2002, 
ApJ, 574, L33 
\reference Bono, G., Marconi, M., \& Stellingwerf, R. F. 1999a, ApJS, 122, 167 
\reference Bono, G., Petroni, S., \& Marconi, M. 2003, in Interplay between 
Periodic, Cyclic and Stochastic Variability in Selected Areas of the HR 
Diagram, ed. C. Sterken (San Francisco: ASP), astro-ph/0212183  
\reference Breger, M. 2002, in Observational Aspects of Pulsating B- and 
A Stars, ed. C. Sterken \& D.W. Kurtz (San Francisco: ASP), 17  
\reference Capaccioli, M., Cappellaro, E., Della Valle, M., D'Onofrio, M.,
Rosino, L., \& Turatto, M. 1990, ApJ, 350, 110  
\reference Caputo, F., Castellani, V., Marconi, M., \& Ripepi, V. 2000, 
MNRAS, 316, 819  
\reference Caputo, F., Marconi, M., Musella, I., \& Pont, F. 2001, A\&A, 
372, 544 
\reference Castellani, V. 1999, in Globular Clusters, ed. C. Martinez Roger,
I. Perez Fournon, \& F. Sanchez, (Cambridge, Cambridge Univ. Press), 109 
\reference Clement, C.M. et al. 2001, AJ, 122, 2587  
\reference Dambis, A.K., \& Rastorguev, A.S. 2001, AstL, 27, 108  
\reference Feast, M. 1999, PASP, 111, 775  
\reference Freedman, W. L., HST Key Project 2001, ApJ, 553, 47  
\reference Gregg, M. D. 1997, New Astr, 1, 363  
\reference Groenewegen, M. A. T. 2000, A\&A, 363, 901  
\reference Groenewegen, M. A. T., \& Oudmaijer, R. D. 2000, A\&A, 356, 849  
\reference Han, M., \& Mould, J. R. 1992, ApJ, 396, 453  
\reference Helmi, A. 2002, Ap\&SS, 281, 351  
\reference Helmi, A., \& White, S. D. M. 1999, MNRAS, 307, 495   
\reference Ibata, R., Irwin, M., Lewis, G.F., \& Stolte, A. 2001, ApJ, 547, L133
\reference Ivezic, Z., SDSS, 2000, AJ, 120, 963   
\reference Jensen, J. B., Tonry, J. L., \& Luppino, G. A. 1999, ApJ, 510, 71  
\reference Jerjen, H., \& Tammann, G. A. 1993, A\&A, 276, 1  
\reference  Kaiser, N.; Pan-STARRS Team, 2002,  AAS, 201, 122.07 
\reference Kavelaars, J. J., Harris, W. E., Hanes, D. A., Hesser, J. E.,
 \& Pritchet, C. J.  2000, ApJ, 533, 125  
\reference Kelson, D. D., HST Key Project 2000, ApJ, 529, 768  
\reference Kennicutt, R. C. Jr., HST Key Project 1998, ApJ, 498, 181 
\reference Layden, A.C., Hanson, R.B., Hawley, S.L., Klemola, A.R., \&  
Hanley, C.J.  1996, AJ, 112, 2110 
\reference Leon, S., Meylan, G., \& Combes, F. 2000, A\&A, 359, 907  
\reference Liu, M. C., \& Graham, J. R. 2001, ApJ, 557, L31   
\reference Longmore, A. J., Dixon, R., Skillen, I., Jameson, R. F., 
\& Fernley, J. A.  1990, MNRAS, 247, 684  
\reference Macri, L. M. et al. 2001, ApJ, 559, 243 
\reference Martinez-Delgado, D., Aparicio, A., Gomez-Flechoso, M. A., 
\& Carrera, R.  2001, ApJ, 549, L199  
\reference Odenkirchen, M., Grebel, E.K., Dehnen, W., Rix, H.-W., \& 
Cudworth, K.M. 2002, AJ, 124, 1497  
\reference Portinari, L., \& Chiosi, C.  2000, A\&A, 355, 929  
\reference Saha, A., Sandage, A., Tammann, G. A., Dolphin, A. E.,
Christensen, J., Panagia, N., Macchetto, F. D.  2001, ApJ, 562, 314  
\reference Sandage, A., Bell, R. A., \& Tripicco, M. J. 1999, ApJ, 522, 250 
\reference Sasselov, D. D., et al. 1997,  A\&A, 324, 471  
\reference Suntzeff, N. B., Kraft, R. P., \& Kinman, T. D. 1994, ApJS, 93, 271 
\reference Tammann, G. A., Sandage, A., \& Reindl, B. 1999, in 19th Texas 
Symposium on Relativistic Astrophysics and Cosmology, astro-ph/9904360  
\reference Tanvir, N. R. 1999, in Post-Hipparcos Cosmic Candles, eds.  
A. Heck \& F. Caputo (Dordrecht: Kluwer), 17   
\reference Thomsen, B., Baum, W.A., Hammergren, M., \& Worthey, G. 1997, 
ApJ, 483, L37 
\reference Twarog, B.A., Ashman, K.M., \& Anthony-Twarog, B.J. 1997, AJ, 
114, 2556  
\reference Vivas, A. K., QUEST collaboration, 2001, ApJ, 554, L33  
\reference Walker, A.R. 2000, in IAU Colloq. 176, The Impact of Large-Scale 
Surveys on Pulsating Star Research, ed. L. Szabados \& D.W. Kurtz, 
(San Francisco: ASP), 165     
\reference Watanabe, M., Yasuda, N., Itoh, N., Ichikawa, T., \& 
Yanagisawa, K. 2001, ApJ, 555, 215 
\reference Yanny, B., SDSS,  2000, ApJ, 540, 825    
\end{references}
\end{document}